\documentclass[pra,letterpaper,aps,10pt,superscriptaddress,twocolumn,floatfix,showpacs]{revtex4-1}
\usepackage{graphicx}
\usepackage{float}
\usepackage[section]{placeins}
\usepackage{amsmath}
\usepackage{amsfonts}
\usepackage{amssymb}
\usepackage{epsfig}
\usepackage{epstopdf}
\DeclareGraphicsExtensions{.pdf,.eps,.png,.jpg,.mps}
\usepackage[pdftex]{color}
\usepackage{amsmath,graphicx,amssymb,braket,xcolor,subfigure,upgreek}
\usepackage[colorlinks, linkcolor=blue, citecolor=blue, urlcolor=blue, breaklinks=true]{hyperref}
\usepackage{microtype}
\usepackage{bbm}
\usepackage{color}

\newcommand{\fref}[1]{Fig.~\ref{#1}}
\newcommand{\secref}[1]{Sec.~\ref{#1}}
\newcommand{\ts}[1]{_\text{#1}}

\renewcommand{\Re}[1]{\text{Re}\left\{#1\right\}}
\renewcommand{\Im}[1]{\text{Im}\left\{#1\right\}}

\begin{document}

\title{Polarization Control of Radiation and Energy Flow in Dipole-Coupled Nanorings}
\author{J.~Cremer}
\affiliation{Institut f\"ur Theoretische Physik, Universit\"at Innsbruck, Technikerstr. 21a, A-6020 Innsbruck, Austria}
\author{D.~Plankensteiner}
\affiliation{Institut f\"ur Theoretische Physik, Universit\"at Innsbruck, Technikerstr. 21a, A-6020 Innsbruck, Austria}
\author{M.~Moreno-Cardoner}
\affiliation{Institut f\"ur Theoretische Physik, Universit\"at Innsbruck, Technikerstr. 21a, A-6020 Innsbruck, Austria}
\author{L.~Ostermann}
\affiliation{Institut f\"ur Theoretische Physik, Universit\"at Innsbruck, Technikerstr. 21a, A-6020 Innsbruck, Austria}
\author{H.~Ritsch}
\affiliation{Institut f\"ur Theoretische Physik, Universit\"at Innsbruck, Technikerstr. 21a, A-6020 Innsbruck, Austria}
\date{\today}

\begin{abstract}
Collective optical excitations in dipole-coupled nanorings of sub-wavelength spaced quantum emitters exhibit extreme sub-radiance and field confinement facilitating an efficient and low-loss ring-to-ring energy transfer. We show that energy shifts, radiative lifetimes, and emission patterns of excitons and biexcitons in such rings can be tailored via the orientation of the individual dipoles. Tilting the polarization from perpendicular to tangential to the ring dramatically changes the lifetime of the symmetric exciton state from superradiance to subradiance with the radiated field acquiring orbital angular momentum. At a magic tilt angle all excitons are degenerate and the transport fidelity between two rings exhibits a minimum. Further simulations suggest that, for certain parameters, the decay decreases double-exponentially with the emitter's density. Disorder in the rings' structure can even enhance radiative lifetimes. The transport efficiency strongly depends on polarization and size, which we demonstrate by simulating a bio-inspired example of two rings with $9$ and $16$ dipoles as found in biological light harvesting complexes (LHC). The field distribution in the most superradiant state in a full LHC multi-ring structure shows tight sub-wavelength field confinement in the central ring, while long-lived subradiant states store energy in the outer rings.
\end{abstract}



\maketitle

\section{Introduction}
When several optical dipoles are confined within a region smaller than a wavelength, their radiation properties exhibit strong interactions. Besides the well-known effect of Dicke superradiance~\cite{dicke1954coherence}, other collective eigenstates featuring strong spatial field confinement and sub-radiant properties appear~\cite{asenjo2017exponential,guerin2016subradiance}. Remarkably, a great deal of the extraordinary subradiant properties survive in more extended configurations as long as the distance of neighboring atoms stays below half a wavelength. As a striking example, an infinite chain of dipoles, where each is separated by less than half a wavelength from its neighbors, constitutes a perfectly lossless waveguide for a single photon~\cite{zoubi2010metastability}. For any chain of finite length we still find at least a cubic increase of single-excitation lifetimes with the number of emitters~\cite{plankensteiner2015selective,ostermann2014protected, asenjo2017exponential, kornovan2019extremely}.

In this context regular polygons constitute a particularly interesting geometry, exhibiting an exponential increase of excited state lifetimes with the number of edges~\cite{asenjo2017exponential,davidpaper}. While such configurations, which can be viewed as a minimal instance of a ring resonator, are not so easy to implement experimentally using individual atoms in optical traps, closely related ring-shaped structures of dipoles appear naturally in biological light harvesting complexes~\cite{cogdell2004review,cogdell2006architecture,worster2019structure,brown2019,hu, herek} or can be set up using quantum dot micro-arrays~\cite{richner2016printable}. Alternatively, one could study such structures in tweezer arrays employing a transition with a correspondingly larger wavelength~\cite{barredo2016arrays2D,endres2016array,barredo2018arrays3D} or microwave setups based on superconducting qubits~\cite{das2017interfacing}.

Experimentally, clear signatures of subradiance in dense atomic clouds were first observed in 2016~\cite{guerin2016subradiance}. More recently, collective back scattering from uniformly filled optical lattices~\cite{bettles2016enhanced,Shahmoon2017mirror,manzoni2018optimization} has been observed experimentally~\cite{bloch2020mirror}. Aside from studying the surprising and exotic properties of dark states in such coupled dipole arrays~\cite{davidpaper} it was recently suggested that they could also serve as emitters and antennas for single photon or exciton transfer with low loss and high fidelity~\cite{davidpaper,asenjo2017exponential}.
\begin{figure}[b]
\centering
 \includegraphics[width=0.7\columnwidth]{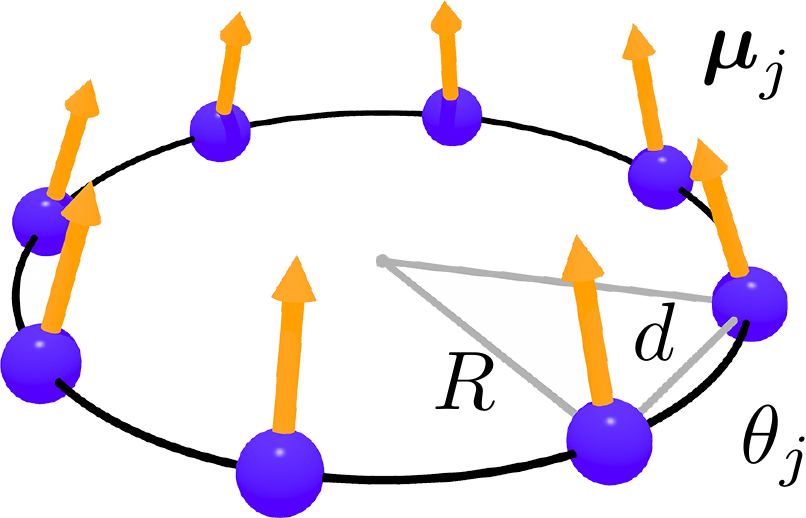}
 \caption{\emph{Ring (regular polygon) of quantum emitters.} The emitters are separated by a sub-wavelength distance $d$ on a ring of radius $R$ and feature prescribed dipole orientations $ \boldsymbol{\mu}_j$, indicated by yellow arrows. The angular separation between two neighboring dipoles is denoted by $\theta_j$. In our specific examples we restrict ourselves to rotationally symmetric orientations.}
 \label{fig:model}
\end{figure}

In this work, we study generalized single- and multi-ring structures in more detail. Starting from the effect of changing the dipole orientation with respect to the ring's plane which allows for controlling their special properties, we also simulate the effect of small disorder in single rings. Inspired by biological examples in light harvesting complexes we investigate the coupling between rings of substantially different size and calculate the special properties of multi-ring configurations. Due to the complexity of the full system, we put our emphasis on the field distribution and the nonlinear effects arising from multiple excitations in such configurations. In this context, the term exciton (biexciton) refers to a (two) delocalized electronic excitation(s) within the structure.

First, we define our mathematical model and review the central properties of excitons as a function of the ring's size and the number of dipoles. In \secref{sec:radiation} we demonstrate how the properties of the energy shifts and lifetimes in the single excitation manifold can be controlled via the dipoles' orientations, where central properties transpire to the spatial radiation pattern. New and surprising scaling behaviors of the radiative lifetimes with the atom number and their distances are discussed in the next section, followed by an extension to multiple excitations in the ring. \secref{sec:transport} is dedicated to the effective coupling of quantum states between rings of equal and different size. In the final section we highlight remarkable properties of field distributions in coupled multi-ring ensembles, inspired by the biological structures of efficient photosynthesis complexes.

\section{Model}
\subsection{Dipole-Dipole Interactions}
We consider an ensemble of $N$ identical two-level atoms with ground state $\ket{g}$ and excited state $\ket{e}$ at fixed positions $\mathbf{r}_i$ for $i=1,...,N$ in free space. The two states are energetically separated by $\omega_0$ and feature an electric transition with a dipole moment $\boldsymbol{\mu}_i$ with identical modulus, $\left| \boldsymbol{\mu}_i \right| = \mu$, for all atoms. The corresponding lowering operator acting on the $i$-th atom is denoted by $\sigma_i$. The inherent coupling to the surrounding vacuum modes leads to an effective dipole-dipole interaction between each pair of atoms. These interactions can be viewed as an interference of the fields emerging from each individual atom. The spontaneously emitted field from the atomic ensemble is given by~\cite{asenjo2017exponential}
\begin{equation}
\textbf{E}^+(\textbf{r},t) = \frac{3\Gamma_0}{4\mu^2}\sum_j \textbf{G}(\textbf{r}-\textbf{r}_j,k_0)\cdot\boldsymbol{\mu}_j\sigma_j(t).
\end{equation}
Here, $\Gamma_0 = \mu^2 k_0^3 / 3\pi \epsilon_0$ (note that we set $\hbar=1$) is the rate of spontaneous emission from a single atom, its natural linewidth, and $k_0 = \omega_0/c$ is the wavenumber associated with the atomic transition frequency. The field propagator $\textbf{G}$ is the Green's tensor of an oscillating unit dipole source in free space, i.e.\
\begin{equation}
\begin{aligned}
\textbf{G}(\textbf{r},k_0) &= e^{ik_0 r} \left[ \left(\frac{1}{k_0 r} + \frac{i}{(k_0r)^2} - \frac{1}{(k_0r)^3}\right)\mathbbm{1} \right. \\
&- \left. \frac{\textbf{r}\textbf{r}^\text{T}}{r^2}\left(\frac{1}{k_0 r} + \frac{3i}{(k_0 r)^2} - \frac{3}{(k_0 r)^3}\right)\right].
\end{aligned}
\end{equation}
From this, the atomic dynamics can be split into coherent and incoherent collective processes. On the one hand, the coherent part of the dipole-dipole interaction $\Omega_{ij}$ between atoms $i$ and $j$ is incorporated into the Hamiltonian as
\begin{equation}
H\ts{dip} = \sum_{i,j:i\neq j} \Omega_{ij} \, \sigma_i^\dag\sigma_j.
\end{equation}
The incoherent part of the dipole-dipole interaction, on the other hand, is accounted for by a Lindblad term describing collective decay,
\begin{equation}
\mathcal{L} \left[ \rho \right] = \frac{1}{2}\sum_{ij}\Gamma_{ij}\left(2\sigma_j\rho\sigma_i^\dag - \sigma_i^\dag\sigma_j\rho - \rho\sigma_i^\dag\sigma_j\right).
\end{equation}
Here, $\Gamma_{ij}$ is the mutual decay rate of atoms $i$ and $j$, and $\Gamma_{ii}=\Gamma_0$. The collective rates are given by the real and imaginary part of the of the overlap of the respective dipole moments with the Green's tensor,
\begin{subequations}
\label{Eq:coupling}
\begin{align}
\Omega_{ij} &= -\frac{3\Gamma_0}{4\mu^2}\Re{\boldsymbol{\mu}_i^*\cdot \textbf{G}(\textbf{r}_i-\textbf{r}_j,k_0)\cdot\boldsymbol{\mu}_j},
\\
\Gamma_{ij} &= \frac{3\Gamma_0}{2\mu^2}\Im{\boldsymbol{\mu}_i^*\cdot \textbf{G}(\textbf{r}_i-\textbf{r}_j,k_0)\cdot\boldsymbol{\mu}_j}.
\end{align}
\end{subequations}
The system dynamics are then described by the master equation,
\begin{equation}
\dot{\rho} = i \left[ \rho,H\ts{dip} \right] + \mathcal{L} \left[ \rho \right].
\end{equation}

If we restrict our investigations to the single excitation manifold, the dynamics of the system are captured by an effective non-Hermitian Hamiltonian,
\begin{align}
H\ts{eff} &= \sum_{i,j}\left(\Omega_{ij} - i\frac{\Gamma_{ij}}{2}\right)\sigma_i^\dag\sigma_j.
\label{eq:Heff}
\end{align}
The complex eigenvalues $\lambda_m$ of $H_{\mathrm{eff}}$ incorporate the collective frequency shifts and decay rates of the eigenstates of the system, i.e.\
\begin{subequations}
\begin{align}
J_m &= \Re{\lambda_m},\\
\Gamma_m &= -2 \Im{\lambda_m}.
\label{eq:gamma_m}
\end{align}
\end{subequations}

\subsection{Excitons in Rotationally Symmetric Ring Configurations}
In the following we consider $N$ two-level atoms uniformly arranged on the edges of a regular polygon (a ring) separated by a distance $d$, as depicted in \fref{fig:model}. We assume the dipole orientations to be arranged rotationally invariant as well. The system then shows special properties due to its high symmetry. In this generic case it is possible to analytically diagonalize the effective Hamiltonian~\eqref{eq:Heff} and find its eigenstates in the form of rotationally invariant spin waves~\cite{davidpaper} as
\begin{equation}
\ket{\psi_{m}} = \frac{1}{\sqrt{N}} \sum_{j} e^{i m \theta_{j}} \ket{e_{j}},
\end{equation}
where $m \in \left \lbrace 0, \pm 1, \pm 2, \ldots \pm \lfloor (N-1)/2 \rfloor \right \rbrace$ is the angular momentum quantum number of the state, $\theta_j = 2 \pi (j-1)/N$, and $\ket{e_{j}} = \sigma_j^\dag \ket{g}^{\otimes N}$. Clearly, each atom has an equal amount of population and thus each eigenstate consists of a fully delocalized excitation. The corresponding eigenvalues are
\begin{equation}
\lambda_m = - \frac{3 \Gamma_0}{4 \mu^2 N } \sum_{j,l} e^{i m (\theta_{l} - \theta_{j})} \boldsymbol{\mu}_j^* \cdot \textbf{G}(\mathbf{r}_j - \mathbf{r}_l, \omega_0) \cdot \boldsymbol{\mu}_l.
\end{equation}
Note, that the spectrum is symmetric, so $\lambda_m$ = $\lambda_{-m}$. Furthermore, the state with $m = 0$ is non-degenerate, whereas the highest momentum modes are doubly-degenerate if the number of emitters $N$ is odd.

The excitations can also be understood as delocalized quasi-particles, excitons, with a momentum $k_m$ given by
\begin{equation}
k_m d = \frac{2\pi m}{N}.
\end{equation}
This interpretation allows for an intuitive explanation of subradiance: excitons with a wavenumber larger than the wave number of the surrounding vacuum $k_0$ can no longer couple to the free-space modes~\cite{asenjo2017exponential}. Thus, spontaneous emission is widely inhibited.

\section{Radiation Properties and Exciton Energy Shifts in Single Rings}
\label{sec:radiation}
For small rings the single excitation states $\ket{\psi_{m}}$, are delocalized and energetically shifted from the bare atomic resonance by many linewidths with spontaneous decay rates spanning orders of magnitude (see \fref{fig:exciton_shifts_lifetimes}). In the following we show that their exotic properties strongly depend on the geometry and, in particular, on the orientation of the dipoles.

In order to limit the complexity of the discussion we focus on the case of rotationally invariant geometries, corresponding to a $C_N$ symmetry for N atoms: typical generic cases are i) all dipoles parallel and perpendicular to the plane of the ring, ii) all oriented tangentially or iii) all pointing in the radial direction. These three cases exhibit a very different radiation behavior and scaling of emission rates and energy flow with size. Extraordinary phenomena appear at special intermediate angles as they are realized in biological light harvesting systems. Due to the presence or lack of mirror symmetries, even and odd numbered polygons show a qualitatively different spectrum with a unique maximally dark state appearing for even particle numbers only.

\subsection{Energy Shifts and Lifetimes as a Function of the Dipole Orientation}
Looking at the coupling strengths in~\eqref{Eq:coupling} one can expect energy shifts and lifetimes of excitons to be correlated. In particular, the relative dipole orientations determine whether the bright or the dark modes are higher in energy: when all polarization vectors are parallel and oriented transversely to the plane of the ring, we observe predominantly repulsive interactions at short distances and thus the symmetric bright modes are strongly shifted upwards, whereas for the tangential case, when dipoles point approximately towards each other, we obtain downward shifts of the bright modes.

\begin{figure}[h]
\centering
\includegraphics[width=\columnwidth]{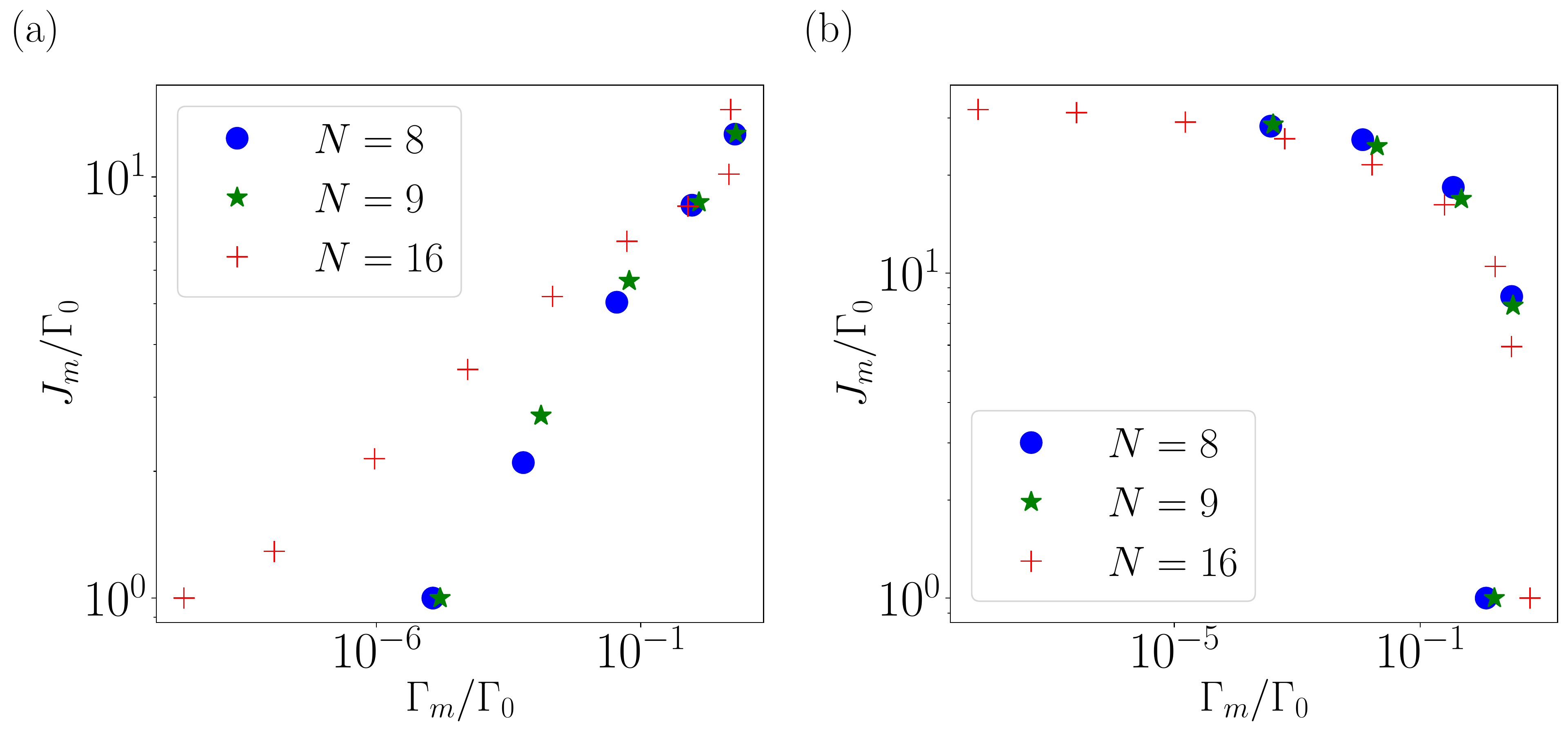}
\caption{\emph{Polarization dependence of energy shifts.} Energy shift $J_m$ of the $m$-th eigenstate vs.\ its decay rate $\Gamma_m$ in the single-excitation manifold at an inter-particle distance $d = 0.1 \lambda_0$ for atom numbers $N \in \lbrace 8,9,16 \rbrace$. We see strong but opposite correlations for (a) transverse and (b) tangential polarization of the dipoles.}
\label{fig:exciton_shifts_lifetimes} 
\end{figure}

We depict this strong correlation at an atom distance $d=0.1 \lambda_0$ comparing different atom numbers in \fref{fig:exciton_shifts_lifetimes}. The lifetimes vary over several orders of magnitude, while energy shifts span a bit more than one magnitude only, showing a strong correlation or anti-correlation, respectively. Note the double degeneracy of the most dark state for odd atom numbers $(N=9$).

\begin{figure}[b]
\centering
\includegraphics[width=\columnwidth]{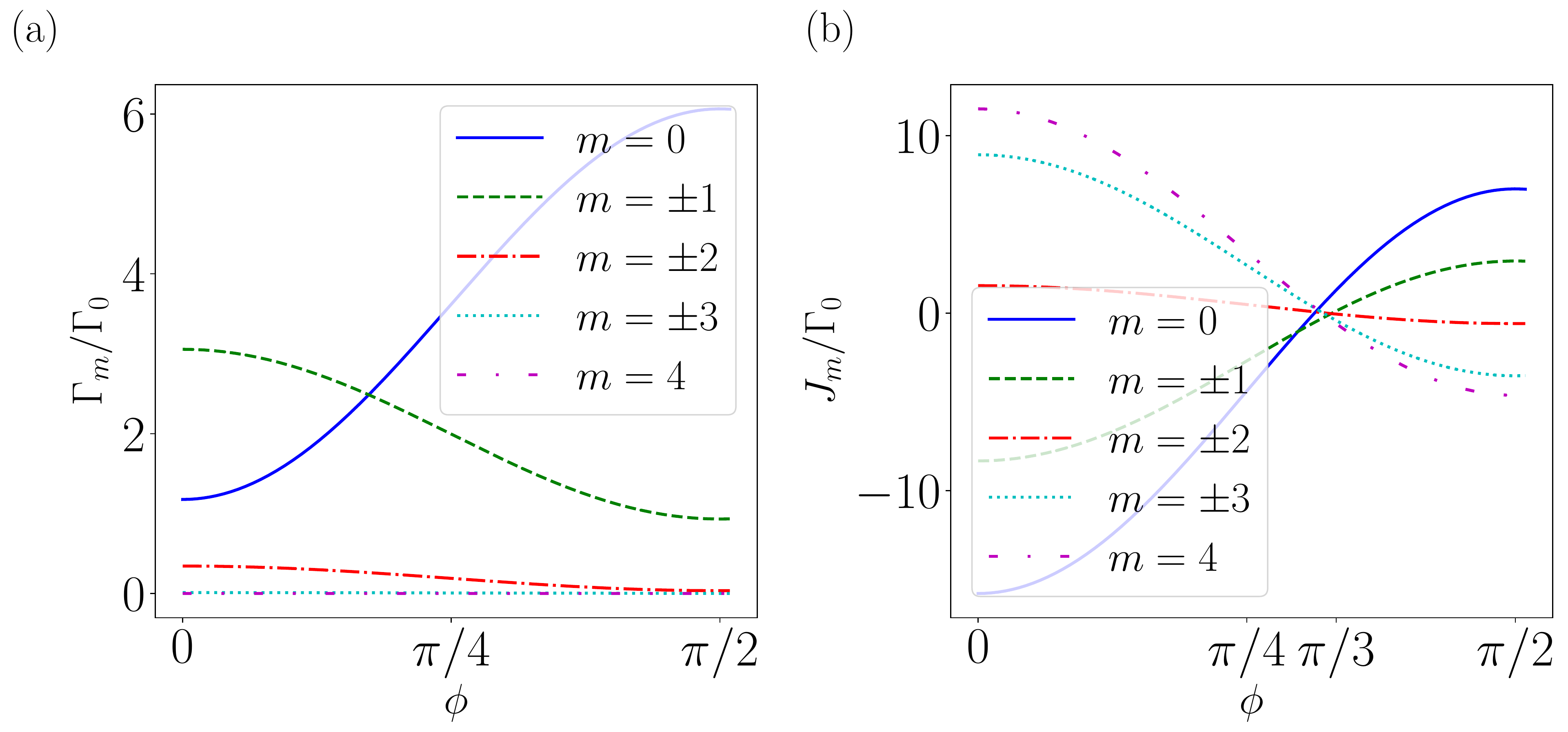}
\caption{\emph{Decay rates and energy shifts for different modes.} (a) Decay rates $\Gamma_m$ and (b) Energy shifts $J_m$ for modes $m \in \lbrace 0,\pm 1,\pm 2,\pm 3, 4 \rbrace$ in a ring with $N = 8$ as a function of the dipole angle $\phi$. The angles $\phi = 0$ and $\phi = \pi/2$ correspond to tangential and transverse atomic polarizations, respectively. For tangential polarization two equivalent bright modes occur for $m = \pm 1$, whereas for transverse polarization only one bright mode for $m = 0$ exists. The mode $m = 4$ is the most sub-radiant one in both cases ($d/\lambda_0 = 0.1$).}
\label{fig:angular_shifts_lifetimes}
\end{figure}

The above two limiting cases lead to the question how these two qualitatively different energy-lifetime correlations change, when we continuously rotate the polarization from one configuration to the other. To study this we calculate the collective decay rates as a function of the dipole orientation for a ring with $N = 8$ emitters. We start with a tangential orientation ($\phi = 0$) and rotate our dipoles upwards ending up at a transverse polarization ($\phi = \pi/2$). In \fref{fig:angular_shifts_lifetimes}a we can see an increase of the decay rate of the symmetric mode ($m = 0$), while the higher order modes $m=\pm 1$ decay more slowly, and those with $m = \pm 2$ become subradiant. Moreover, we find that the modes with the highest angular momentum $m = \pm 3, 4$ are highly subradiant, regardless of the polarization angle $\phi$. In \fref{fig:angular_shifts_lifetimes}b we show the corresponding energy shifts. Changing from tangential to transverse polarization the energy shifts change sign and their absolute values decrease significantly. Interestingly, we observe that at a certain value of $\phi$ all mode energy shifts become approximately equal to zero. This finding is analysed in more detail in the next sub-section. 

\begin{figure}[t]
\centering
\includegraphics[width=\columnwidth]{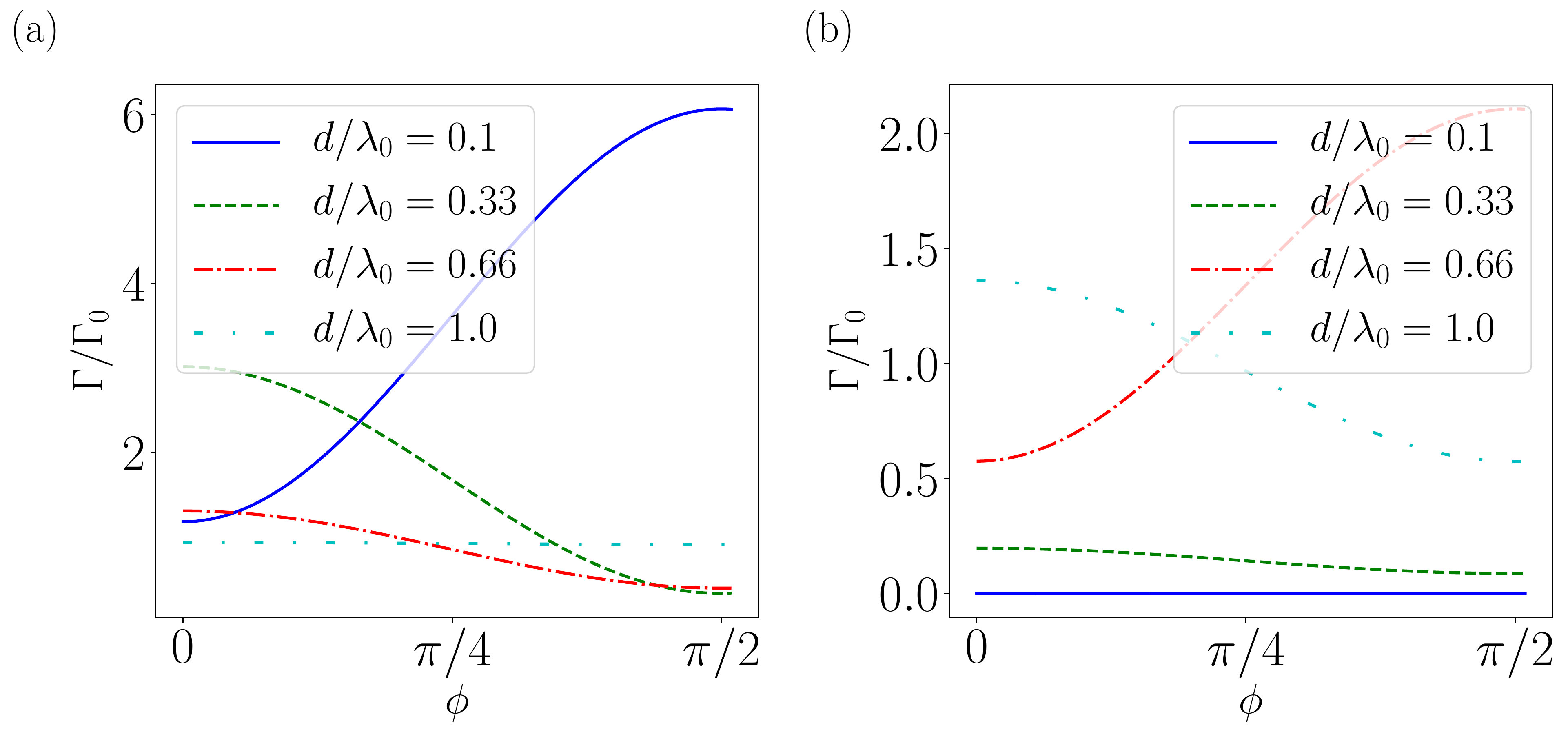}
\caption{\emph{Angular dependence of decay rates.} Polarization angle dependence of the decay rates $\Gamma_m$ for (a) the symmetric mode m $= 0$ and (b) the highest eigenmode $m = 4$ in a ring with $N=8$ for different distances $d/\lambda_0$.}
\label{fig:angular_rates}
\end{figure}

This particular behavior also depends on the ring's size as shown in \fref{fig:angular_rates}, where we plot the collective decay rates of the symmetric ($m=0$) and subradiant ($m=4$) modes of a $N=8$ atom ring, as a function of the dipole orientation $\phi$ for different inter-particle distances. We find that the symmetric mode, which has a maximal decay rate for dense and transversally polarized atomic rings, becomes maximally radiant for tangential polarization instead if the interatomic distance becomes large enough. Moreover, the mode with the largest angular momentum is subradiant for both cases with $d/\lambda_0 < 1/2$ and it becomes darker as the polarization rotates towards $\pi/2$.

In summary, we observe that for transverse polarization the lowest momentum modes at small distances yield the most pronounced superradiance, while the large momentum modes exhibit strong subradiance independent of the dipole orientation. Comparing tangential and transverse polarization, we find that the latter shows much stronger collective effects in the super- as well as the subradiant regime.

\begin{figure}[t]
\centering
\includegraphics[width=\columnwidth]{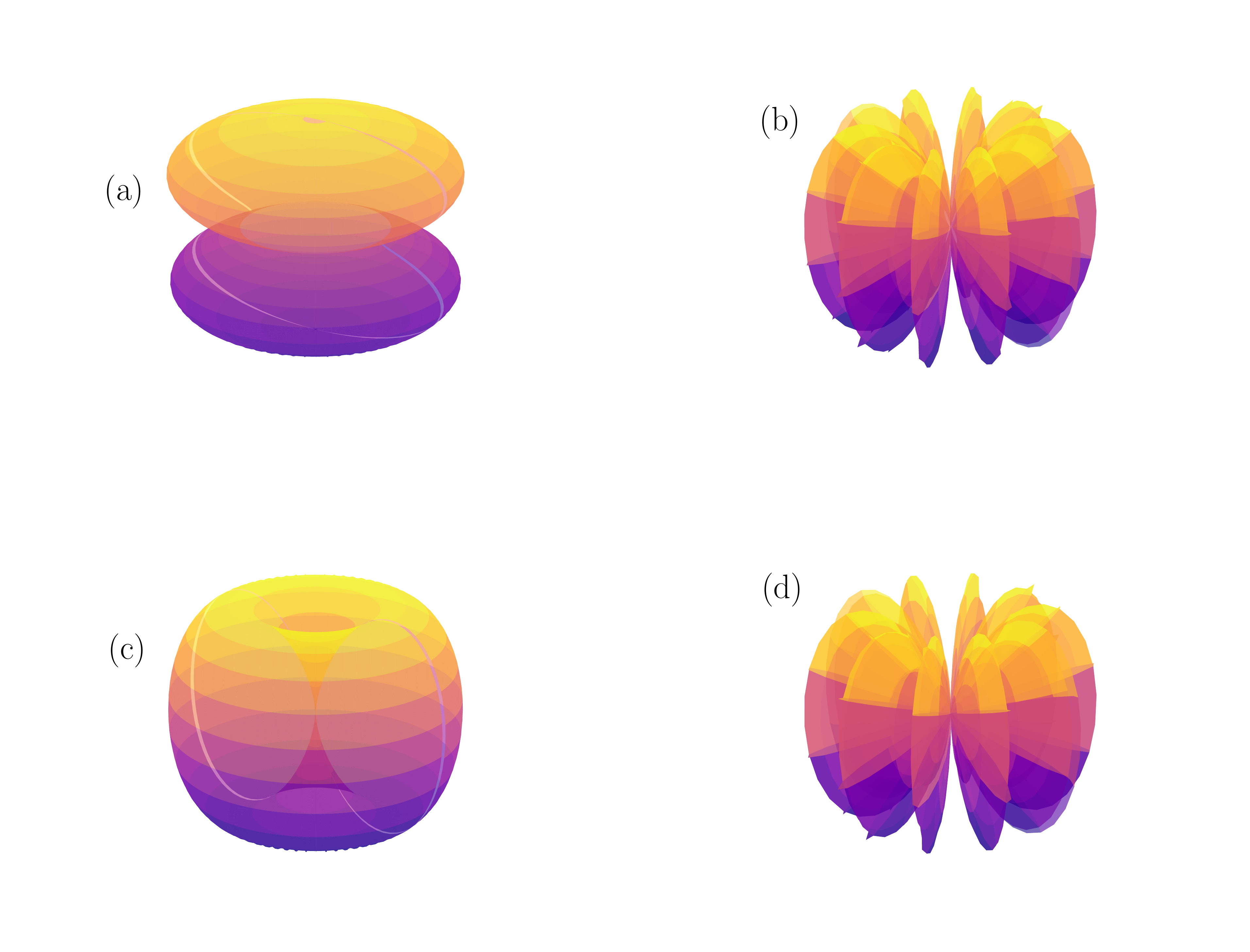}
\caption{\emph{Far-field radiation patterns.} Intensity distribution of the radiated far field at a distance of $200 R$ from the center of a ring of $N = 10$ atoms with $d/\lambda_0 = 0.1$. (a), (b) Tangential polarization, (c), (d) transverse polarization. The left-hand side column, (a) and (c), corresponds to the most superradiant states ($m=0$ and $m=\pm 1$, respectively) while the right-hand side column, (b) and (d), shows the most subradiant state, i.e.\ m=5. Note, that the purpose here is to highlight the distinct spatial character of the respective fields rather than a quantitative comparison, which is why the field magnitudes are shown on arbitrary (unequal) scales. The color is used for better visibility.}
\label{fig:single_ring_radiation_patterns}
\end{figure}

This behavior has important consequences for the far-field radiation patterns, which we show for tangential as well as transverse polarization in \fref{fig:single_ring_radiation_patterns}. For tangential polarization we can see that in the superradiant regime the field is strongly transverse and in the subradiant regime the far field is transversally evanescent.

\subsection{"Magic" Dipole Orientation}
In the previous section we have found that the energy shifts of all collective modes of a rotationally symmetric ring nearly vanish at a particular dipole orientation. Analytically, it can be proven that in the limit of a very dense ring ($d/\lambda \to 0$ and $N\to \infty$) there exists a dipole orientation for which the energy shifts of all collective modes are exactly zero. In this limit, the dipole-dipole interaction $\Omega_{ij}$ between atoms $i$ and $j$ in~\eqref{Eq:coupling} reduces to its short-range contribution,
\begin{equation}
\Omega_{ij} \to \frac{-3\Gamma_0}{4\mu^2 k_0^3 r_{ij}^3} \left[ \left( \boldsymbol{\mu}^*_i \cdot \hat{\mathbf{r}}_{ij} \right) \left( \boldsymbol{\mu}_j \cdot \hat{\mathbf{r}}_{ij} \right) - \boldsymbol{\mu}^*_i \cdot \boldsymbol{\mu}_j \right].
\end{equation}
We proceed by parametrizing our dipoles in cylindrical coordinates as
\begin{equation}
\boldsymbol{\mu}_i = \mu \cos \phi \left( \alpha \hat e_{r,i} + \beta \hat e_{t,i} \right) + \mu \sin \phi \hat e_z
\end{equation}
with the local basis $\left( \hat e_{r,i}, \hat e_{t,i}, \hat e_z \right)$ and $|\alpha|^2 + |\beta|^2 = 1$. For a ring with polarizations preserving rotational symmetry, $\alpha$, $\beta$ and $\phi$ necessarily assume the same values for each dipole in the ring. For a mode with an angular momentum $m$, after summing up the energies over all pairs and exploiting the symmetry, we obtain
\begin{equation}
J_m = \sum_{j \not = \ell} e^{i m(\theta_\ell -\theta_j)} \Omega_{\ell j} = N \sum_{j=1}^{N-1} e^{-i m \theta_j} \Omega_{Nj},
\end{equation}
with
\begin{equation}
\Omega_{N j} = \frac{-3\Gamma_0 \, \left[ \cos^2 (\phi) \left( 3 |\beta|^2 - \sin \left( \theta_j /2 \right) \right) -1 \right]}{32 [k_0 r \sin (\theta_j/2)]^3}.
\end{equation}

Striving for a total energy shift of zero ($J_m = 0$) in the the limit of large $N$, this reduces to
\begin{equation}
\cos\phi = \sqrt{\frac{1}{3 |\beta|^2}},
\end{equation}
which is independent of the value of $m$. For tangential polarization ($\beta = 1$) this corresponds to $ \phi = \cos^{-1}(1/\sqrt{3}) \approx 0.953 \sim 54.7^\circ$. Note, that in this limit the magic dipole orientation can exist for $|\beta| > \sqrt{1/3}$ only.

\subsection{Radiation Properties of a Ring with Disorder}
\begin{figure}[h]
\centering
\includegraphics[width=\columnwidth]{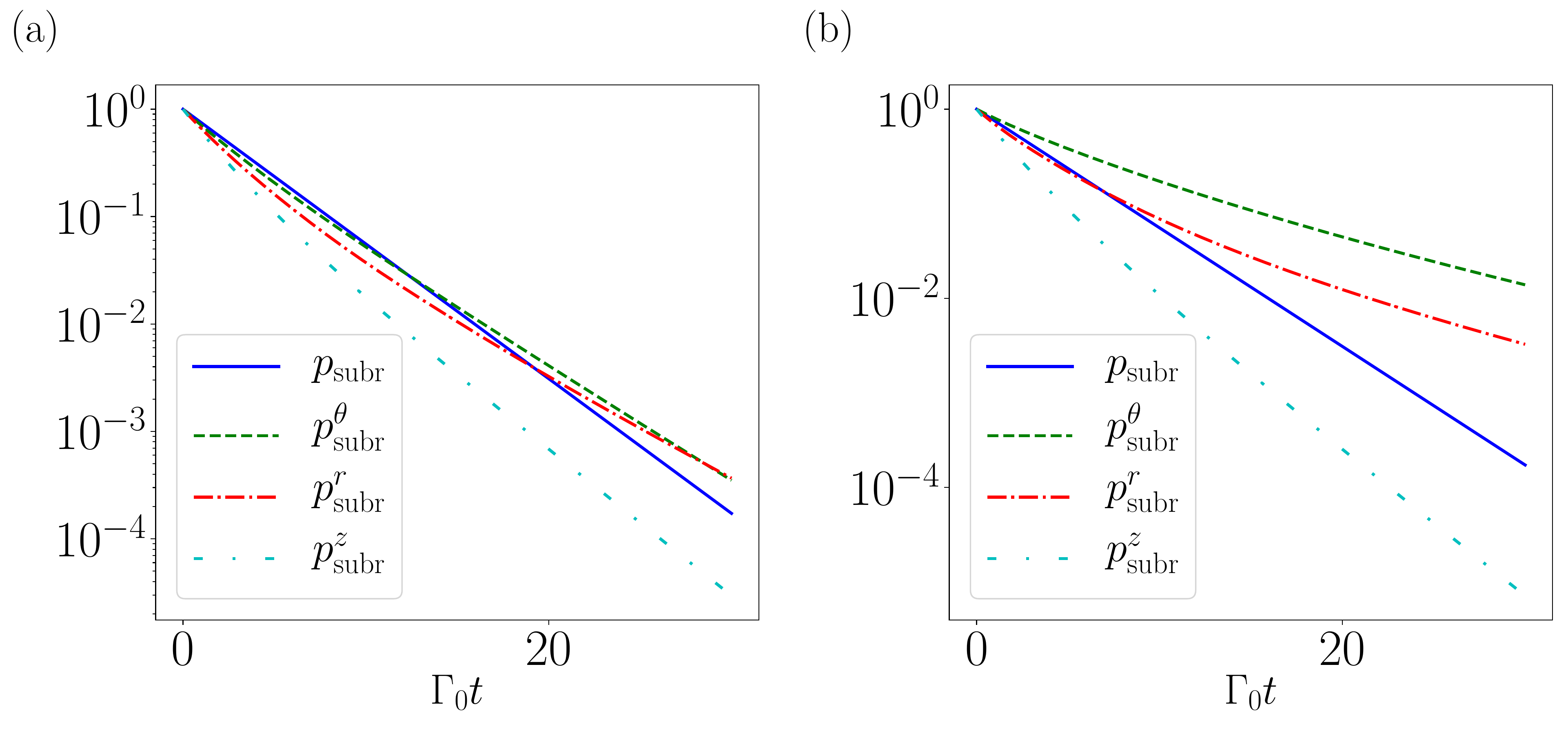}
\caption{\emph{Robustness of subradiance in the presence of disorder.} Excited state population decay averaged over $100$ random realizations of disorder of different type (angular, radial or vertical shifts) as indicated in the legend for a ring with $N=8$ emitters and average distance $d/\lambda_0 = 0.4$. For each realization we choose the most subradiant state. (a) Random displacements up to $d/5$ and (b) up to $0.4 d$. For random radial displacements of up to $40\%$ a significantly more subradiant mode appears when compared to the perfectly regular case.}
\label{fig:disorder_decay}
\end{figure}

Let us now investigate the impact of imperfect geometries by adding a small spatial disorder in the ring's atomic positions. For this we allow for three types of random displacements: (i) moving the atoms along the ring, while maintaining the ring shape, (ii) radially displacing the emitters and (iii) moving the emitters out of the plane of the ring. For each random configuration we find the most subradiant state and let it evolve under the disordered Hamiltonian. We then average over the evolved excited state population for $100$ realizations. The result (denoted respectively by $p_\text{subr}^\theta$, $p_\text{subr}^r$ and $p_\text{subr}^z$ for the three types of disorder) is shown in \fref{fig:disorder_decay} for a ring with an average interatomic distance $d/\lambda_0 = 0.4$. As reference, we also plot the decay of the unperturbed state denoted by $p\ts{subr}$.

In perfect rings the sub-radiant state is extremely stable. Upon introducing position disorder we observe that this stability is initially reduced, so that for large disorder the state decays several orders of magnitude faster at the beginning, see \fref{fig:disorder_decay}a. However, after quite some time it appears that the lifetime is even increased due to disorder. One exception here is when the emitters are moved out of the plane of the ring. Then, the decay is always enhanced. For larger disorder, the effects we find are even more pronounced, resulting in a remarkable reduction of the decay rates when disordering the atoms along the ring preserving its shape, see \fref{fig:disorder_decay}b. Note that this enhancement of subradiance with disorder is also present in chains~\cite{plankensteiner2015selective}.

\section{Scaling of Subradiance with Ring Size and Atom Number}
In this section we will investigate the scaling behavior of sub-radiance in more detail, as the basis of further investigations below. We concentrate on the most sub-radiant decay channel as a function of the atom number $N$ and inter-particle distance $d/\lambda_0$. It was previously shown that for $d=\lambda_0/3$ the decay rate of the most subradiant state in a ring scales down exponentially with the number of atoms~\cite{asenjo2017exponential,davidpaper}. Recent related investigations of regular chains of emitters have shown, that for a single excitation a strong polynomial reduction with up to $N^{-6}$ of the most subradiant decay rate can occur at certain distances~\cite{kornovan2019extremely}. For two excitations in a chain of atoms coupled to a waveguide, it was found that there are states which feature an even lower decay rate than the most subradiant state of the single-excitation manifold~\cite{zhang2019subradiant}.

\subsection{Single Excitation Subradiance}
\begin{figure}[t]
\centering
\includegraphics[width=\columnwidth]{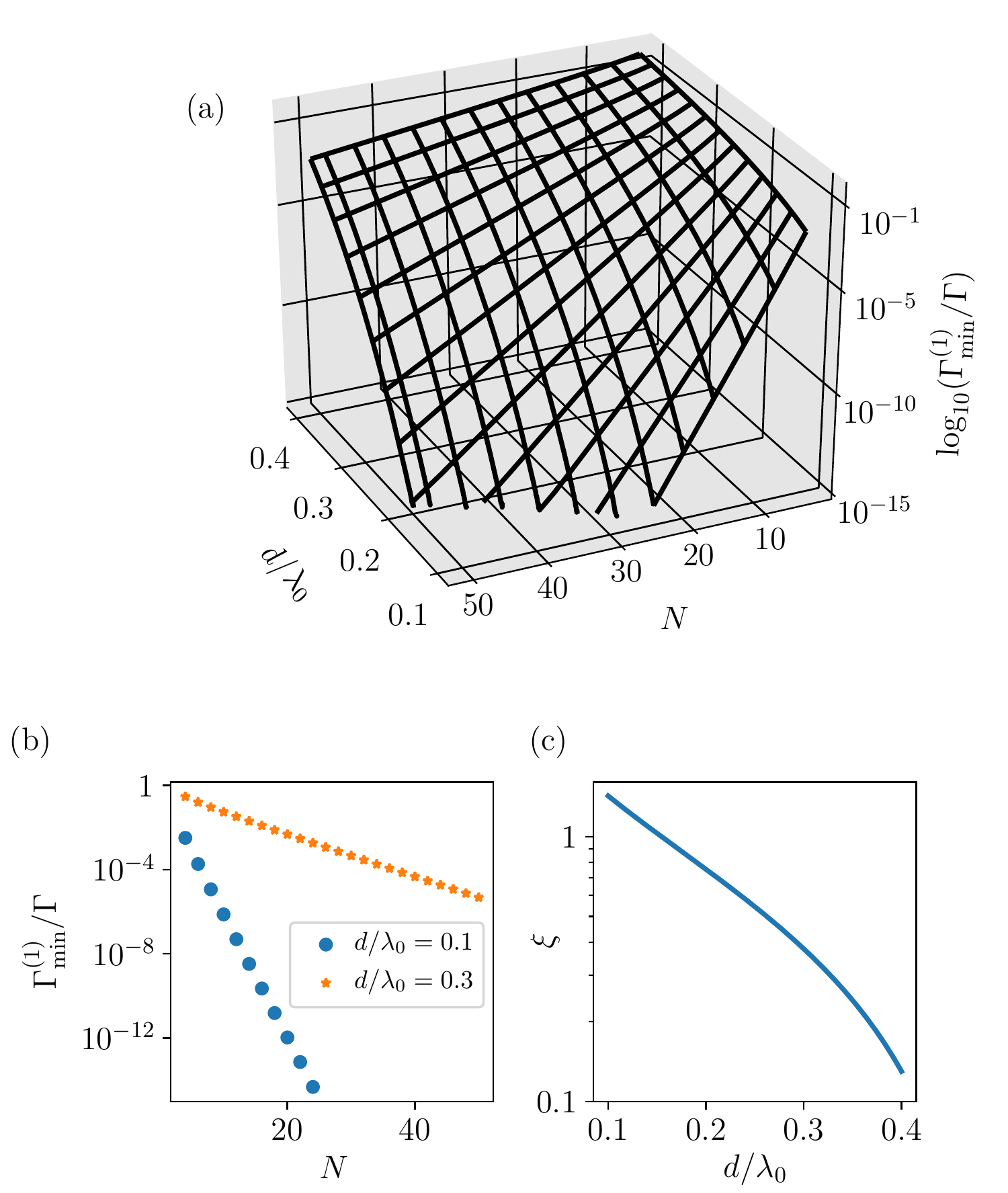}
\caption{\emph{Scaling of the minimal decay rate in the single-excitation manifold.} (a) The minimal decay rate in the single-excitation manifold, $\Gamma_\mathrm{min}^{(1)}$ as a function of $N$ and $d/\lambda_0$. Because of the logarithmic scale for the decay rate, the linear behavior with $N$ hints at an exponential reduction. (b) This exponential scaling with $N$ is even stronger for smaller distances. (c) The exponent $\xi$ is shown as a function of the distance. For $d/\lambda_0 \lesssim 0.3$ we observe a close to exponential scaling of $\xi$ with $d/\lambda_0$. The polarizations were chosen to be orthogonal to the plane of the ring ($\phi = \pi/2$). In all graphs we cut off our data at $\Gamma_\mathrm{min}^{(1)}< 10^{-15}\Gamma_0$ in order to avoid numerical errors.}
\label{fig:scaling_single}
\end{figure}

First, we investigate the behavior of the most subradiant state featuring a single excitation only. The decay rate is computed as in~\eqref{eq:gamma_m} for the largest possible $m$, namely
\begin{equation}
\Gamma\ts{min}^{(1)} = \Gamma_{\lfloor N/2\rfloor}.
\end{equation}
In \fref{fig:scaling_single}a, we show this decay rate as a function of both, the number of atoms $N$ and the inter-particle spacing $d/\lambda_0$. Clearly, the largest suppression of decay occurs at small separations and a large number of atoms, i.e.\ at maximal density. While the reduction of the decay rate with decreasing $d/\lambda_0$ exhibits polynomial scaling, we find that for all considered distances the decay reduces exponentially with $N$. Yet, we do not only observe an exponential reduction with $N$, but also an increase in the absolute value of its exponent with smaller separations. This can also be seen in \fref{fig:scaling_single}b, where we plot the scaling with $N$ for two different distances. We can therefore propose that the most subradiant decay rate scales as
\begin{equation}
\Gamma\ts{min}^{(1)}\propto \exp \left( -\xi N \right),
\end{equation}
where the exponent $\xi$ itself is a function of the distance $d/\lambda_0$.

The scaling of $\xi$ with distance is depicted in \fref{fig:scaling_single}c. On a logarithmic scale $\xi$ shows a linear scaling for $d/\lambda_0 \lesssim 0.3$. Therefore, the exponent itself scales exponentially at small distances. This suggests that at distances below $\lambda_0/3$, the most subradiant decay rate of a single excitation is suppressed in a double-exponential manner with increasing atomic density.

Overall, these results emphasize how extremely subradiant a nano-ring of dipole-coupled atoms can become. The corresponding states for sufficiently large $N$ and at distances $d\ll\lambda_0$ are almost perfectly decoupled from the environment and therefore extraordinarily long-lived.

\subsection{Subradiance with Two Excitations}
\begin{figure}[b]
\centering
\includegraphics[width=\columnwidth]{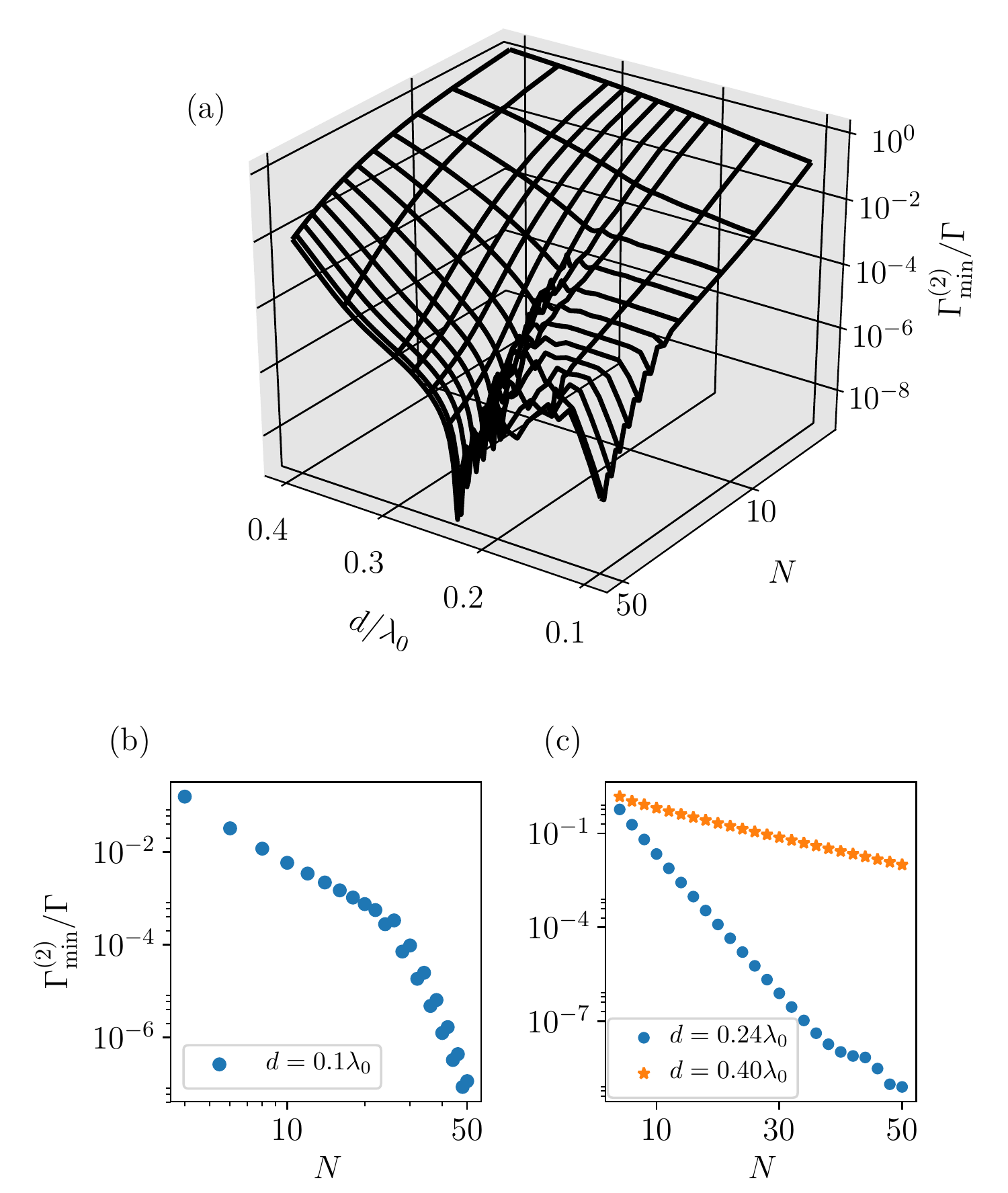}
\caption{\emph{Scaling of the minimal decay rate in the two-excitation manifold.} (a) The minimal decay rate for states in the two-excitation manifold is shown as a function of the distance $d/\lambda_0$ and the atom number $N$. (b) The decay rate at the smallest considered distance scales polynomially in $N$, which can be seen from the linear behavior on the log-log scale. At $N\sim 30$, the polynomial suppression increases abruptly. (c) For larger distances, we still find an exponential suppression of the decay with growing system size $N$. At the distance featuring the overall lowest decay rate from (a), no clear scaling law can be identified for all $N$.}
\label{fig:scaling_double}
\end{figure}
Most investigations so far have been based on the single excitation regime, i.e. the weak excitation case, which allows for an analytical assessment, while still being practically relevant and physically very interesting. For more than one excitation the effective Hamiltonian, in general, cannot be diagonalized analytically anymore. Since the number of states in the $k$-excitation manifold $\binom{N}{k}$ grows rapidly, we restrict ourselves to $k=2$ in order to limit the numerical effort.

In \fref{fig:scaling_double}a we show the behavior of the minimal decay rate involving two excitations, $\Gamma\ts{min}^{(2)}$, as a function of the number of atoms and the particle spacing. The overall situation is less clear than for a single excitation, since, depending on the distance $d/\lambda_0$, distinctly different scalings in $N$ can be found. At small separations we observe a polynomial scaling of the decay rate with $N$. This can also be seen in \fref{fig:scaling_double}b, where it becomes clear that at a certain size of the system, a stronger polynomial suppression of the decay occurs. At larger distances the increase in lifetime still grows in an exponential fashion, as shown in \fref{fig:scaling_double}c.

The fact that we find a polynomial scaling at close distances for more than one excitation can be interpreted as follows: a single excitation within a finite equidistant chain, where the light scatters off the ends of the chain, shows a minimal decay rate which reduces polynomially~\cite{asenjo2017exponential,kornovan2019extremely}. In a ring, an excitation cannot propagate through an already excited atom. An excitation within a ring thus constitutes a defect for a second one. A ring containing two excitations therefore features open boundaries as a chain does for a single excitation, which leads to a similar polynomial suppression of decay rates.

Another feature, which is reminiscent of a chain, is that at a certain inter-particle distance, $d/\lambda_0\approx 0.24$, we find a distinctly lower minimal decay rate. This is similar to the findings of Ref.~\cite{kornovan2019extremely}, where it has been shown that for optimal distances, the polynomial decrease of the decay with $N$ becomes much more drastic than for one excitation. No clear scaling law can be identified for this distance: for relatively small $N$ it appears to be exponential, but for larger systems this is no longer the case.

\section{Excitation Transport between Rings}
\label{sec:transport}
Inspired by the so-called light harvesting complexes (LHC) occurring in biological systems, where a structure consisting of several coupled rings of different molecules was observed~\cite{cogdell2006architecture,cogdell2004review,worster2019structure,hu}, we study the coupling strengths and the energy transport between two coupled rings of different size. Note that in the following considerations the rings are arranged such that they are closest at exactly one site of each ring and their centers lie on the same axis (site-site configuration).

The coupling strength between the two rings prepared in the modes with well defined angular momentum $m_1$ and $m_2$ is given by
\begin{align}
\lambda_{m_1,m_2} &= \frac{1}{N} \sum_{\substack{i\in\mathcal{R}_1,\\ j\in\mathcal{R}_2}} \left(\Omega_{ij} - i\frac{\Gamma_{ij}}{2}\right) e^{i (m_1 \theta_{i}-m_2 \theta_{j} )},
\label{eq:coupling-evals}
\end{align}
where as a shorthand notation, we have defined two sets of indices, one for the sites in the first ring (with $N_1$ emitters), $\mathcal{R}_1=\{1,2,...,N_1\}$, and one for the sites in the second ring (with $N_2$ emitters), $\mathcal{R}_2=\{N_1+1,...,N_1+N_2\}$. The dispersive and dissipative couplings can then be found from $J_{m_1,m_2} = \textrm{Re} \{\lambda_{m_1,m_2}\}$ and $\Gamma_{m_1,m_2} = -2 \textrm{Im} \{\lambda_{m_1,m_2}\}$, respectively. In addition, we define the coupling efficiency between the two modes~\cite{davidpaper} as 
\begin{equation}
\eta_{m_{1}, m_{2}} = \frac{J^{2}_{m_{1}, m_{2}}}{(4 \Delta^{2}_{m_{1}, m_{2}} + \max \{\Gamma^{2}_{m_{1}}, \Gamma^{2}_{m_{2}}\})},
\end{equation}
with $\Delta_{m_1,m_2} = J_{m_1}-J_{m_2}$ being the difference in energy of the two ring modes. 

\begin{figure}[t]
\centering
\includegraphics[width=\columnwidth]{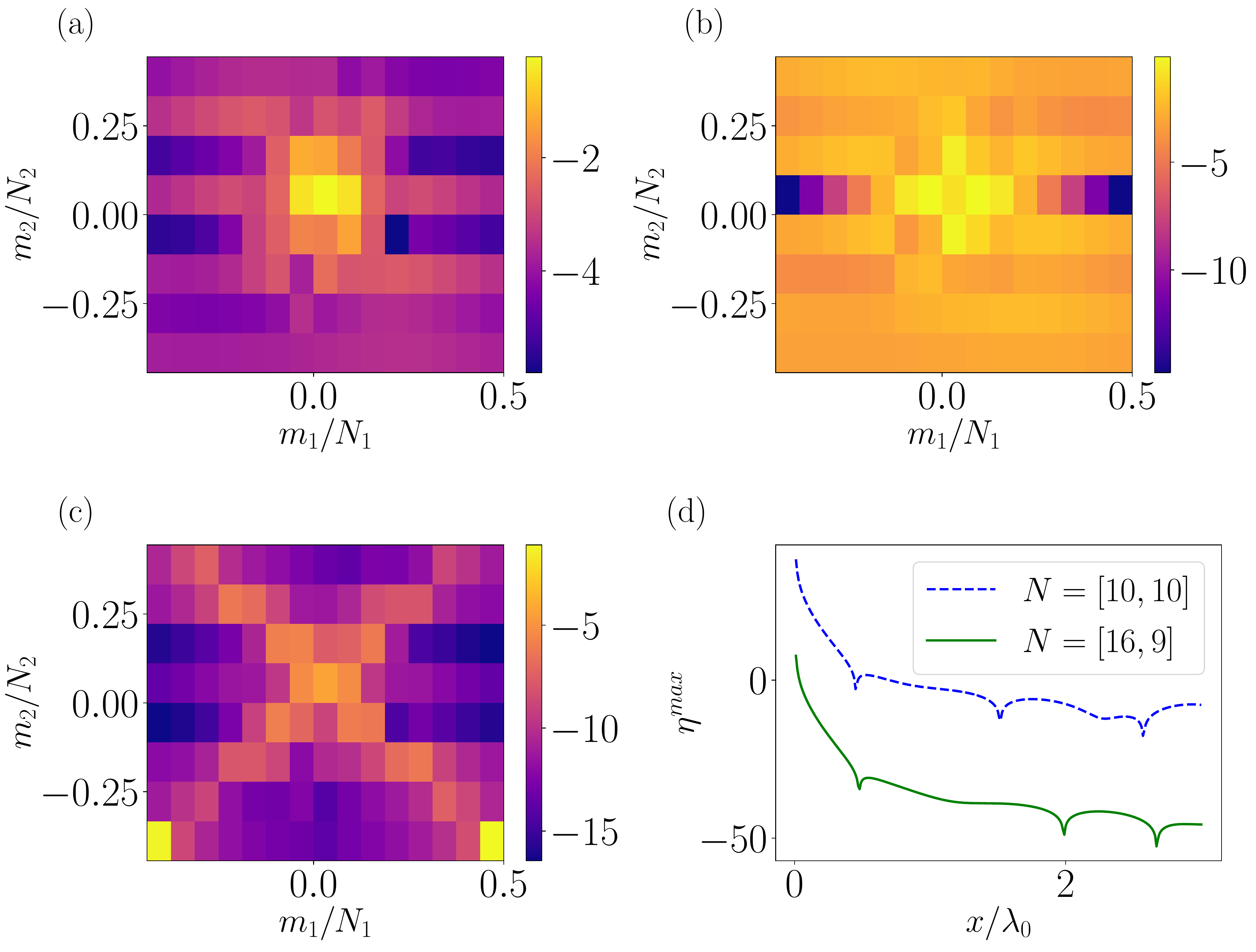}
\caption{\emph{Coupling between eigenmodes of two unequal rings with transverse polarization} Two rings with $N_1=16$ and $N_2=9$ at $d/\lambda_0 = 0.1$, separated by the distance $x = 0.12\lambda_0$. (a) Dispersive coupling $J_{m_1,m_2}$, (b) dissipative coupling $\Gamma_{m_1,m_2}$, (c) coupling efficiency $\eta_{m_1,m_2}$ and (d) maximal coupling efficiency for $m_1 = m_2 = \lfloor N/2\rfloor$ as a function of the ring-to-ring distance $x/\lambda_0$ (green, solid). For comparison, the result for equally sized rings is shown as well (blue, dotted). All the couplings as well as the efficiency are shown on logarithmic scales.}
\label{fig:coupling_two_sizes}
\end{figure}

In \fref{fig:coupling_two_sizes} we evaluate the dispersive and dissipative couplings, as well as the coupling efficiency between two rings with $N_1 = 16$ and $N_2 = 9$ emitters with transverse polarization and separated by a distance $x = 0.12 \lambda_0$, as a function of the angular momentum of the two rings $m_1$ and $m_2$. We find that the dispersive coupling is rather large in the superradiant, but nearly vanishes in the subradiant regime. Thus, the coupling efficiency yields non-negligible values for $m_1$ = $-m_2$ only. However, they are still several magnitudes smaller compared to equally sized rings~\cite{davidpaper}, since for unequal rings $\Delta_{m_1,-m_1}$ is non-zero.

The maximum value of the efficiency $\eta_{m_{1},m_{2}}^{\mathrm{max}} = \eta_{m_{1}=\lfloor N/2\rfloor, m_{2}=\lfloor N/2\rfloor}$ as a function of the ring-to-ring separation $x/\lambda_0$ is shown in \fref{fig:coupling_two_sizes}d. It oscillates, but also decreases exponentially for increasing distances. For comparison, the result of two equally sized rings, which can reach much larger values, is shown as well.

\subsection{Efficient Exciton Transfer between Two Rings}
For equally sized rings, it has been shown that the fidelity of transferring a wave-packet is extraordinarily high and only slowly damps out with time~\cite{davidpaper}. Thus, an excitation is transferred with almost no loss between the rings over a long period of time. In a similar manner we now want to evaluate the fidelity of the energy transfer between two differently sized rings for a Gaussian wave-packet centered at the site $k$ farthest from the second ring, i.e.\
\begin{equation}
\ket{\Psi_{i,k}^{m}} = \frac{1}{\sqrt{n}} \sum_{j \in R_{i}} e^{i \theta_{j} m} e^{- \frac{\vert \vec{r}_{j} - \vec{r}_{k} \vert }{2 R^{2} \Delta \theta^{2}}} \ket{e_{j}},
\label{eq:gauss}
\end{equation}
where $n$ accounts for the normalization, $\Delta \theta$ denotes the angular spread of the wave-packet of width $R \Delta \theta$, and $m$ is the central momentum. For an infinitely wide wave-packet, $R\Delta \theta \to\infty$, the expression in~\eqref{eq:gauss} reduces to an eigenstate of our system with the corresponding angular momentum $m$. For a guided mode in the first ring with momentum $m$, we expect that when the mode travels to the second ring, it will invert its momentum. So, for a finite width wave-packet we can assume that it is transferred to the second ring thereby inverting the momentum to $-m$, but remaining otherwise unchanged. With this, we can put down the fidelity $\mathcal{F}$ of creating this wave-packet in the second ring as
\begin{equation}
\mathcal{F}(t) = \underset{k}{\max} \Big\{ \braket{\Psi_{2,k}^{-m}|\Psi (t)} \Big\}.
\label{eq:fidelity} 
\end{equation}
Here, $\ket{\Psi (t)}$ is given by the time evolution in the truncated Hilbert space and the initial condition $\ket{\Psi (0)}$ = $\ket{\Psi_{1,k}^{m}}$, and the maximization over the site $k$ is required due to the fact that we cannot predict the position of the wave-packet created in the second ring at all times.

\begin{figure}[t]
\centering
\includegraphics[width=\columnwidth]{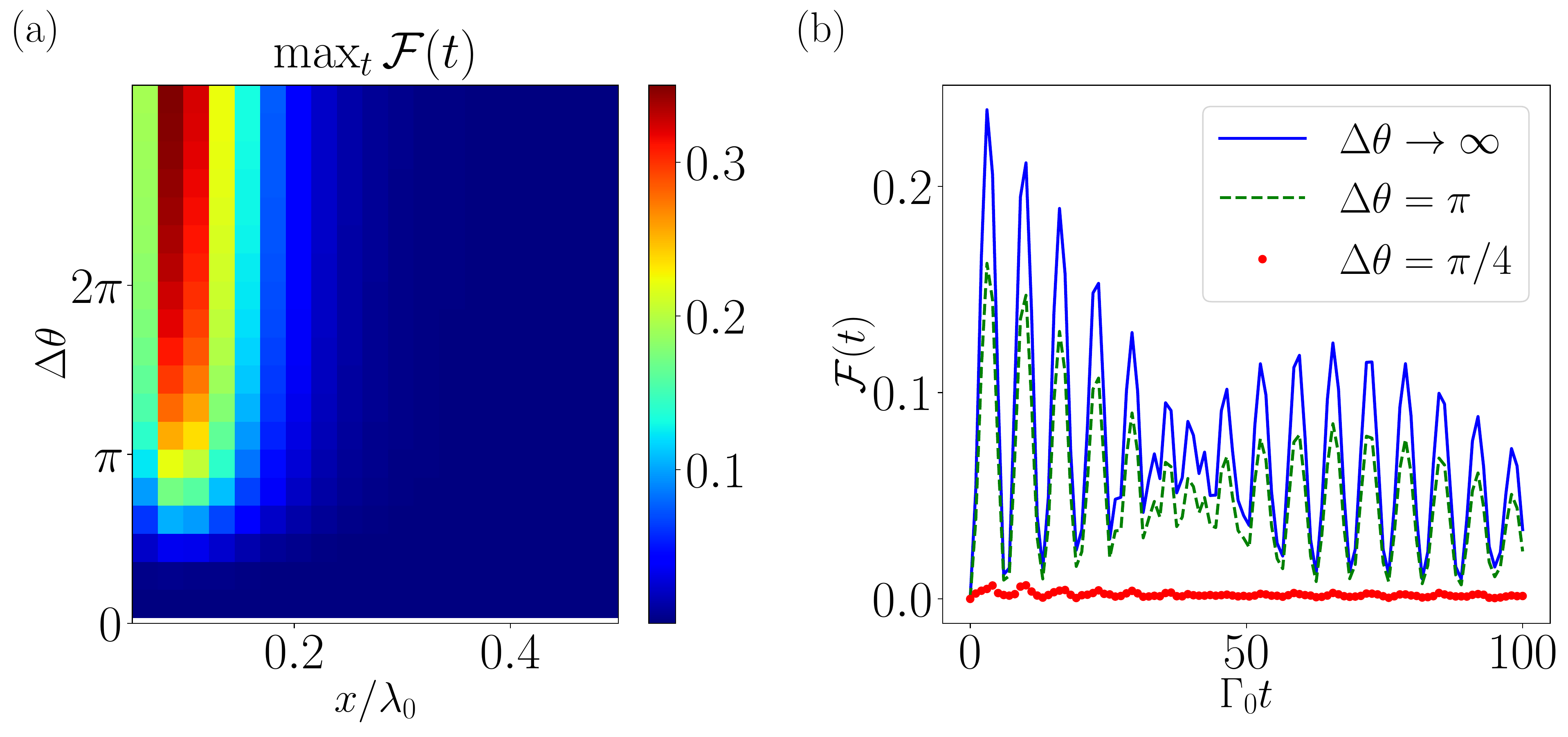}
\caption{\emph{Fidelity of wave-packet transport between two unequal rings.} Two rings with $N_1 = 9$, $N_2 = 16$ atoms, $d = 0.1\lambda_0$, $m_1=3$, and tangential polarization. (a) Maximal fidelity over time as a function of the ring-to-ring separation $x/\lambda_0$ and the width of the wave-packet $\Delta \theta$. (b) Fidelity as a function of time for a distance $x = 0.12\lambda_0$.}
\label{fig:fidelity_wavepacket_two_sizes}
\end{figure}

In \fref{fig:fidelity_wavepacket_two_sizes}a the maximal fidelity over time as a function of the ring-to-ring separation $x/\lambda_0$ and the width of the wave-packet $\Delta \theta$ is shown. For two rings with a different number of emitters, e.g.\ $N_1= 9$ and $N_2 = 16$, we find the largest fidelities at $x = 0.10 \lambda_0$ to $x = 0.15\lambda_0$ and at a width of $\Delta \theta > 2\pi$. The fidelity is much smaller than for two equally sized rings as in Ref.~\cite{davidpaper}. The time evolution of the exciton transfer is shown in \fref{fig:fidelity_wavepacket_two_sizes}b for two rings separated by $x = 0.12\lambda_0$ and $d = 0.1\lambda_0$. The transport shows a significant decrease on a short time scale followed by an oscillation of the excitation between the rings for a long period with large damping.

\subsection{Exciton Transfer between two rings with "Magic" Dipole Orientation}
So far we have concentrated on the generic cases of vertical or tangential dipole orientation. As we have shown in \secref{sec:radiation}, there is a so called {\sl magic angle} for the dipole orientation in a ring, where all light shifts cancel and all eigenmodes are degenerate. One might ask now, whether this enhances or suppresses energy transport.

\begin{figure}[t]
\centering
\includegraphics[width=\columnwidth]{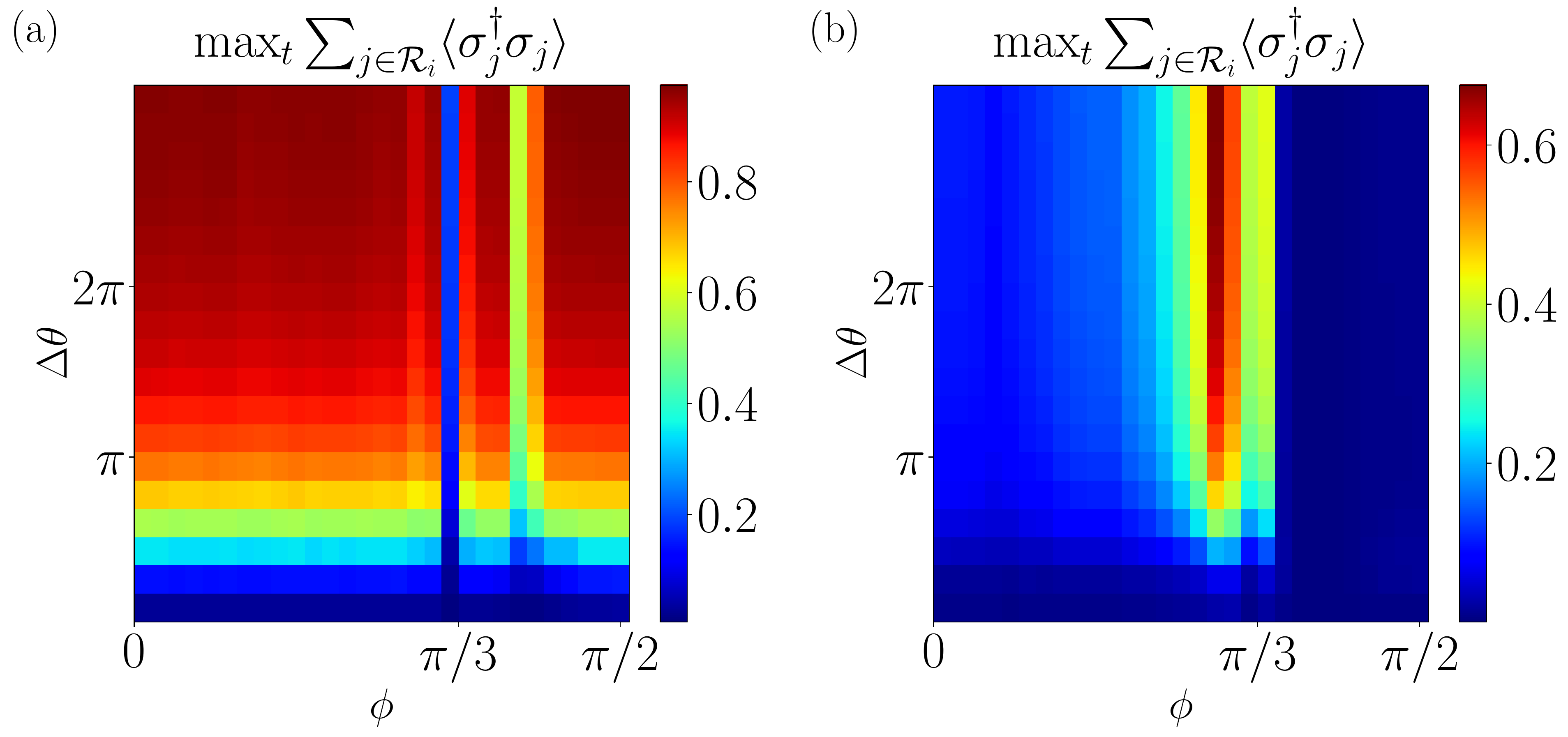}
\caption{\emph{Excitation transfer between two rings.} The maximal excited-state population of the second ring over time is plotted as a function of the dipole orientation $\phi$ and the width of the wave-packet $\Delta \theta$. (a) Equal rings with $N_1=N_2=10$, and $m_1=5$. (b) Unequal rings with $N_1 = 9$, $N_2 = 16$, and $m_1=3$. For unequal rings, the excitation transfer is maximal around the magic orientation $\phi = \cos^{-1}(1/\sqrt{3})$, for which all single ring frequency shifts are almost zero. Contrastingly, for equal rings the population displays a minimum around this point.}
\label{fig:population_angle}
\end{figure}

In \fref{fig:population_angle}a the maximal population in the second ring over time is shown as a function of the dipole orientation $\phi$ and the width of the wave-packet $\Delta \theta$. For two rings with the same number of emitters, e.g.\ $N_1= N_2 = 10$, we find a substantial population transfer nearly independent of the dipole orientation $\phi$. Yet, we observe a significant collapse of the transfer in the region around the magic angle. Since the energy shifts in this area cancel for all eigenmodes, the wave-packet in the first ring couples to multiple different eigenmodes, including radiant ones. This creates an effective loss channel thereby inhibiting the excitation transport.

\begin{figure}[t]
\centering
	\includegraphics[width=\columnwidth]{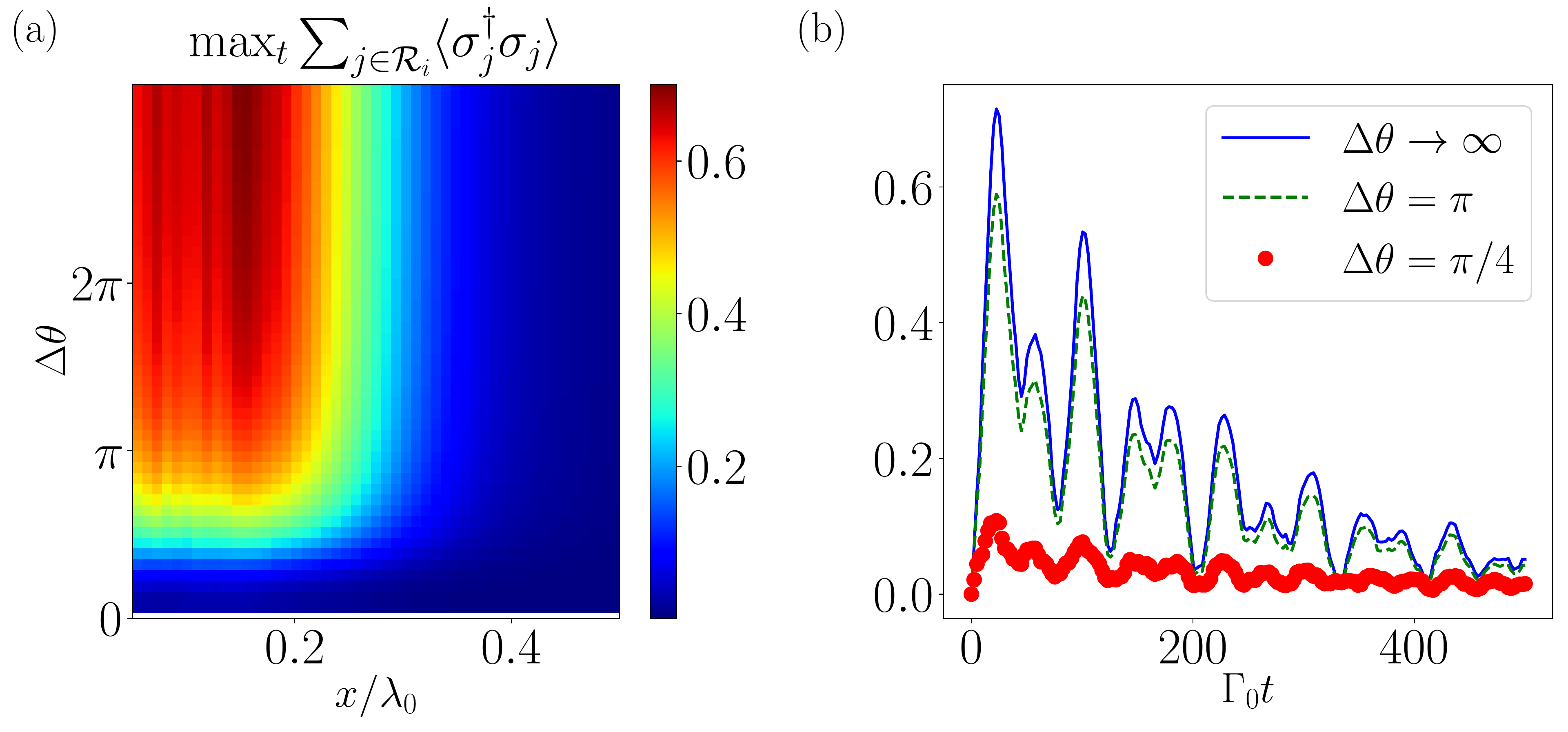}
	\caption{\emph{Excitation transfer between two unequal rings.} Two rings with $N_1 =9$, $N_2=16$, $d/\lambda_0 = 0.1$, $m_1=3$, and magic dipole orientation. (a) The maximal excited state population of the second ring over time as a function of the ring-to-ring separation $x/\lambda_0$ and the width of the wave-packet $\Delta \theta$. (b) Excited-state population of the second ring as a function of time, for a distance $x = 0.17\lambda_0 $.}
		\label{fig:population_magic}
\end{figure}

Comparing these results with the maximal population over time for two differently sized rings, e.g.\ $N_1= 10$ and $N_2 = 16$, as shown in \fref{fig:population_angle}b, we observe the completely opposite behavior: we can achieve a significant increase of the population in the second ring in the region around the magic angle only. If we now compare \fref{fig:fidelity_wavepacket_two_sizes}a and \fref{fig:population_angle}b, we observe that for two differently sized rings the population transfer is remarkably enhanced by choosing the magic dipole orientation.
Consequently, in \fref{fig:population_magic}a the population over time is shown as a function of the ring-to-ring separation $x/\lambda_0$ and the width of the wave-packet $\Delta \theta$. We use the same configuration as in \fref{fig:fidelity_wavepacket_two_sizes}, but at the magic dipole orientation. We find the largest population transfer from $x = 0.15 \lambda_0$ up to $x = 0.18\lambda_0$ and at a width of $\Delta \theta > \pi$.

In \fref{fig:population_magic}b we plot the time evolution of the population transfer. We observe a strong damping after the first run. Interestingly, even for a small width, when the initial state is not an eigenstate of the system and the wave-packet therefore is less localized in momentum space, we can reach a relatively good population transfer.

\begin{figure}[h]
	\centering
	\includegraphics[width=\columnwidth]{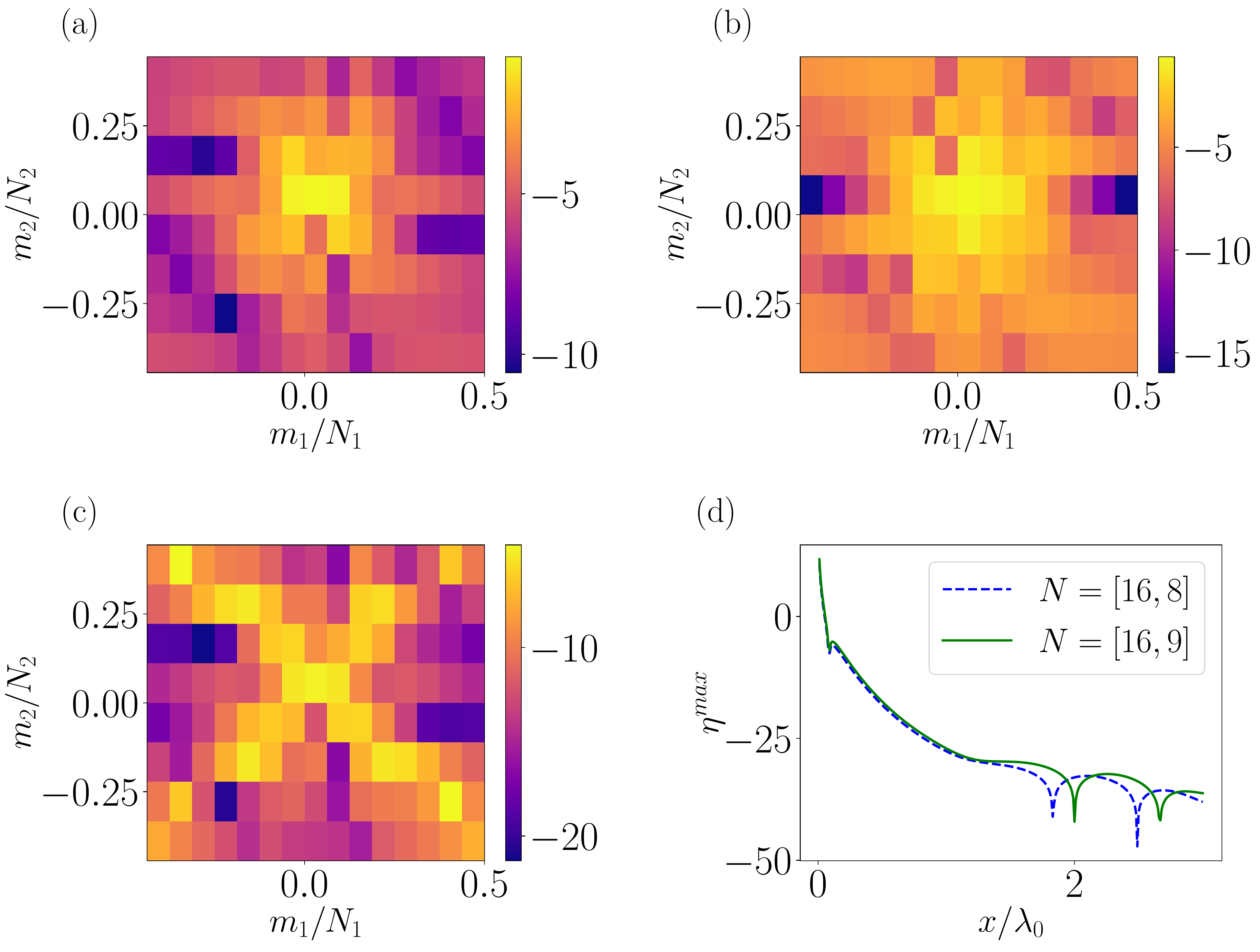}
	\caption{\emph{Coupling between eigenmodes of two unequal rings with magic polarization.} Two rings with $N_1=16$, $N_2=9$, and $d/\lambda_0 = 0.1$, separated by the distance $x = 0.12\lambda_0$ in site-site configuration. (a) Dispersive coupling $J_{m_1,m_2}$, (b) dissipative coupling $\Gamma_{m_1,m_2}$, (c) coupling efficiency $\eta_{m_1,m_2}$ and (d) maximal coupling efficiency for $m_1 = m_2 = \lfloor N/2\rfloor$ as a function of the ring-to-ring distance $x/\lambda_0$ (green, solid). For comparison, the result for rings of different size with an even number of atoms is shown as well (blue, dotted). Note that the respective intensities are on arbitrary scales, so the color scales do not compare quantitatively.}
	\label{fig:couplings_magic_two_sizes} 
\end{figure}

In \fref{fig:couplings_magic_two_sizes} we study the dispersive and the dissipative coupling as well as the coupling efficiency between rings of different size, i.e.\ $N_1 = 16$ and $N_2 = 9$. Both couplings are rather large in the superradiant regions, as one would expect. The coupling efficiency shows an almost diagonal coupling pattern, where, compared to \fref{fig:coupling_two_sizes}c, the sum over all present momenta is a lot larger. This is congruent with what we see in the population transfer.

As shown in \fref{fig:couplings_magic_two_sizes}d, the coupling efficiency is slightly enhanced for the even/odd ($N_1 = 16$, $N_2 = 9$) configuration in contrast to two even numbers of emitters in both rings ($N_1 = 16$, $N_2 = 8$).

\subsection{Collective States of a Bio-Inspired Multiple Ring Configuration}
\begin{figure}[t]
\centering
	\includegraphics[width=\columnwidth]{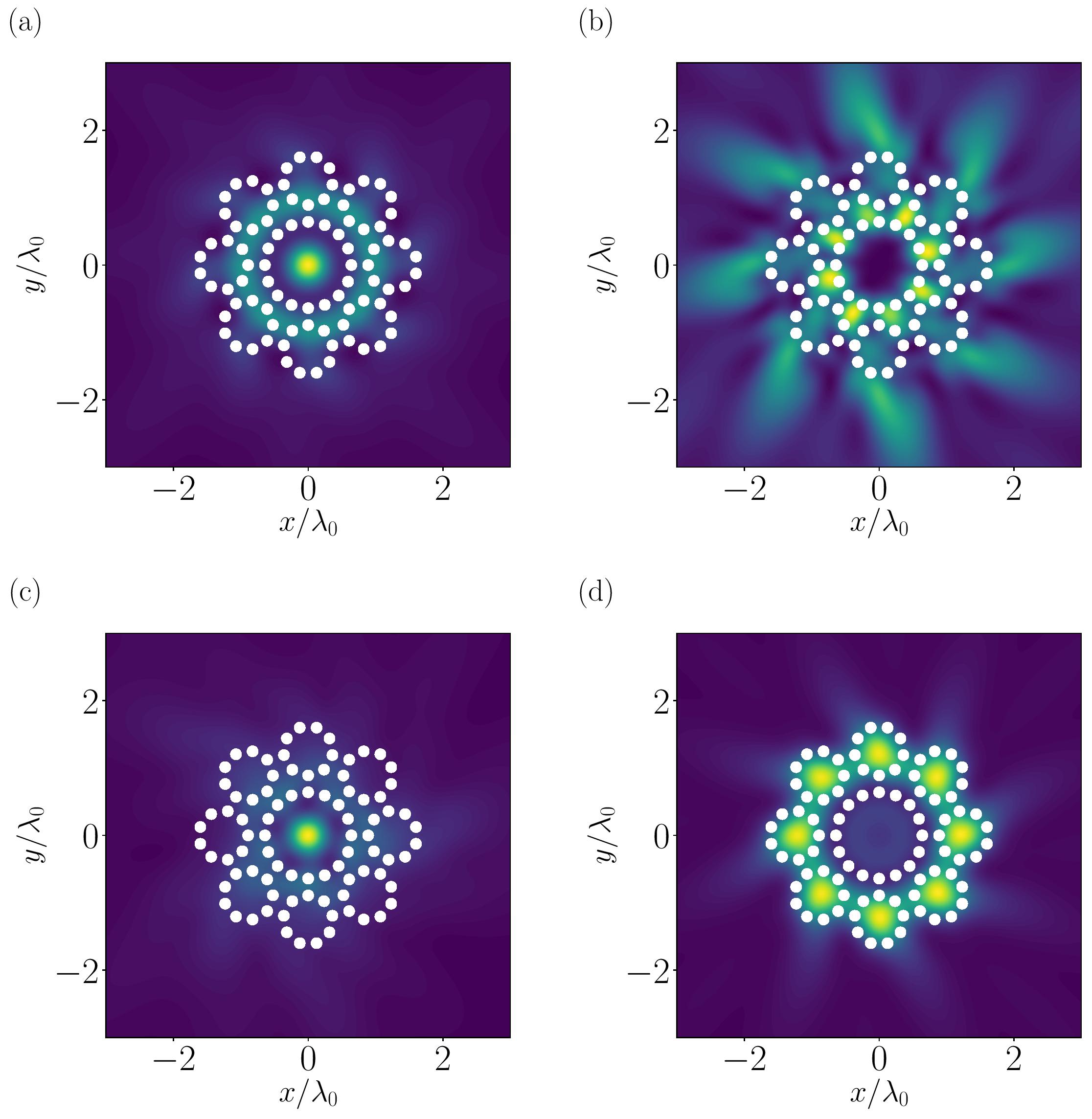}
	\caption{\emph{Field intensity distribution of a bio-inspired multi-ring configuration with tangential polarization.} The central ring has $N\ts{in} = 16$ sites and it is surrounded by eight outer rings each with $N\ts{out} = 9$ emitters. The distance between the atoms in each ring is $d = 0.25\lambda_0$ and equal to the minimal distance between each outer ring and the central one. The field intensity over the plane at $z=2d$ is plotted for a superposition state prepared with (a) outer and central rings in the most superradiant state, (b) outer and central rings in the most subradiant state, (c) outer rings in the most subradiant and central ring in the most superradiant state, and (d) outer rings in the most superradiant and central ring in the most subradiant state.}
	\label{fig:multi_rings_tangential}
\end{figure}
\begin{figure}[t]
\centering
	\includegraphics[width=\columnwidth]{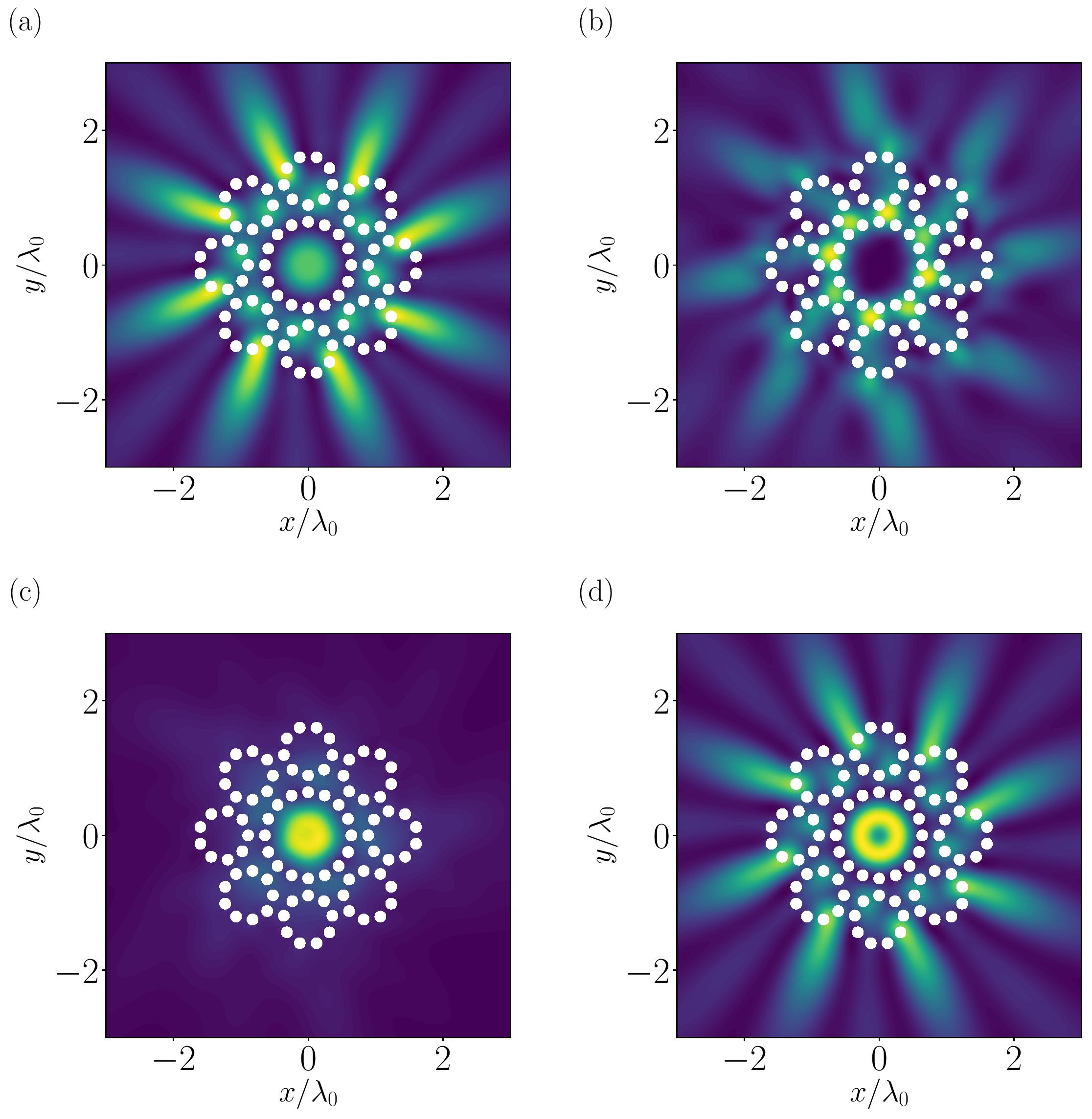}
	\caption{\emph{Field intensity distribution with magic dipole orientation.} The same as in \fref{fig:multi_rings_tangential} but for a dipole moment orientation $\boldsymbol{\mu} /\mu= cos\phi \hat{e}_t + sin\phi \hat{e}_z$, with $\phi = \cos^{-1}(1/\sqrt{3})$.}
	\label{fig:multi_rings_magic}
\end{figure}

The active molecules in different LHCs are arranged in a ring of nine elements. In some bacteria (e.g.\ bacterium Rps. acidophila) eight of these rings are arranged around a central ring of $16$ dipoles containing the reaction center (see Fig. 5 and Fig. 7 in Ref.~\cite{hu}). Here, we will ignore all the complex chemistry as well as phonons and many noise sources, and simply look at a corresponding arrangement of dipole-coupled rings. Interestingly, even this over-simplified model yields some surprising insights already.

In essence, the model reduces to an ensemble of $88$ dipoles in a plane confined within an area much less than $\lambda_0^2$. This area is smaller than the transverse coherence length of any incoming optical photon, which will thus simultaneously couple to all dipoles and excite a bright exciton depending on the light polarization. In \fref{fig:multi_rings_tangential} we depict the intensity of the electric field generated by the state resulting from the superposition of the excitation being in an eigenmode of the central ring and in an eigenmode of each of the outer rings, when all dipoles are oriented along the tangential direction. Specifically, we choose the eigenmodes as the most superradiant or most subradiant states of each of the rings. From \fref{fig:multi_rings_tangential}a and \fref{fig:multi_rings_tangential}c, one can clearly see that, when the central ring is in a superradiant state, the complex has a field maximum in its center. This is where the reaction center would be and thus, the excitation will ultimately be transported there.

Similarly, in \fref{fig:multi_rings_magic} we plot the corresponding results for the magic dipole orientation. On the one hand, we can see that the tilted angles cause significant outward radiation in \fref{fig:multi_rings_magic}a. On the other hand, the field appears to still be focused in the center in \fref{fig:multi_rings_magic}c.

\section{Conclusions}
Nanorings of quantum emitters exhibit extraordinary radiative properties featuring super- and subradiance as well as strong field confinement. Not only are these phenomena prominent in the single excitation manifold but they appear for multiple excitations as well. We have shown that the scaling exponent of the lifetime of the most subradiant states strongly depends on the ring's size, yet, is only marginally altered upon introducing disorder.

Our simulations reveal that many of these properties can be controlled and fine-tuned by adjusting the polarization direction of the individual dipoles. While for transverse dipoles the subradiant states possess the lowest energies and the symmetric superradiant state forms a frequency up-shifted collective dipole, one finds an even qualitatively opposite behavior for a tangential dipole orientation. In the latter case, subradiant states possess higher energy than radiating states and the most symmetric state emits a doughnut-shaped radiation field with orbital angular momentum and zero field in the center. Interestingly, for a dense enough ring, one can find a special dipole orientation, where all single excitation states are almost degenerate. Like individual exciton properties, also ring to ring transfer properties can be tailored with polarization control. While the transfer efficiency is smaller than for equally sized rings, energy transfer between different ring sizes still benefits from subradiance and it can be enhanced by employing the magic polarization. Finally, we have studied the field distributions in complex coupled ring structures as they appear in biological light harvesting systems. Here, dark and bright modes strongly depend on polarization as well. We find that the fields can be confined to the structure's center even stronger than in a single ring as the outer rings act as a sort of isolation layer against environmental decoherence.

\acknowledgments
\noindent We thank Ana Asenjo-Garcia and Claudiu Genes for helpful discussions. We acknowledge funding from the European Union's Horizon 2020 research and innovation program under Grant Agreement No. 820404 iqClock (J.~C., D.~P. and H.~R.), as well as from the Austrian Science Fund under project P29318-N27 (L.~O.). This project has also received funding from the European Union’s Horizon 2020 research and innovation program under the Marie Sk{\l}odowska-Curie grant agreement No. 801110 and the Austrian Federal Ministry of Education, Science and Research (BMBWF). It reflects only the author’s view and the Agency is not responsible for any use that maybe made of the information it contains.

The numerical simulations were performed with the open-source framework QuantumOptics.jl~\cite{Qijulia} and the graphs were produced with the open-source library Matplotlib~\cite{hunter2007matplotlib}.

\end{document}